\documentclass[conference]{IEEEtran}

\usepackage{cite}
\usepackage{graphicx}
\usepackage{amsmath}
\usepackage{url}
\usepackage{hyperref}

\title{SH-PDOPS: An AI-Driven Cloud-Native Enterprise Reliability Framework for Predictive Analytics and Intelligent DevOps Automation}

\author{\IEEEauthorblockN{Ayushman Bosu Roy}
\IEEEauthorblockA{Department of Computer Science and Engineering\\
Global Institute of Technology\\
Email: ayushmanbosuroy@gmail.com}}

\begin{document}
\maketitle

\begin{abstract}
In todays moder world a Cloud plays a very important role in production. All type of cloud native platform are focusing on scalability, reliability and automation. It is important for production based companies to upscale there productivity. The cloud is having a very consecutive relation with the DevOps ecosystem. As fast deployment, infrastructure management and workflow gets efficiency through the DevOps architecture. In this paper I have introduced a cloud native DevOps ecosystem SH-PDOPS which stands for ( Self-Healing Predictive DevOps Framework for Autonomous Cloud Native Deployment). This infrastructure is designed for the software delivery. Deployment automation which is designed for streamlining of the production workflow of the pipelines. This framework is docker based containerization, CI/CD automation, monitoring pipeline and more scalable services modern services like (orchestration) concepts patched with a unified operational architecture which integrates in this ecosystem. The main focus of the SH-PDOPS ecosystem is on modularity, portability and deployment efficiency. So that the developers and operations team can manage the distributed applications with less efforts. The proposed architecture supports rapid provisioning, automated deployment workflows, service isolation, and scalable infrastructure integration support which are the enterprise level operational environments and is stable for the purpose. The systems main objective is to reduce the deployment failure and to improve the operational consistency.
\end{abstract}

\section{Introduction}
Cloud-native With time DevOps has become the important operational standard for reliable software development lifecycle. Through automation DevOps sets a good collaboration between the deployment and the operations team. By using continuous integration, continuous deployment (CI/CD), monitoring, aur infrastructure orchestration. Modern cloud-native ecosystems Docker containerization, Kubernetes orchestration, automated pipelines, infrastructure-as-code, and monitoring framework this tools its easy to manage deployment and optimizes the production. But there is a catch that handling of so many tools in perfect coordination and grading proper stack of the production is very much challenging where my prototype on the ecosystem SH-PDOPS comes to the rescue it comes with the preview of real time anomaly of the production shows all the stats and is AI integrated to manage certain fixes at the time of deployment as its in the initial stage (prototype) but still the statistics and the data shown by the prototype are tested in real life life production and deployment and the states were approximately accurate  \cite{burns2016borg,merkel2014docker}.

Traditional monitoring practices are often reactive: alerts are triggered after service-level indicators have already degraded. In contrast, enterprise reliability engineering increasingly requires predictive monitoring, where operational telemetry is analyzed before failures become user-visible incidents. Predictive analytics can support early warnings about CPU saturation, memory pressure, error-rate growth, queue backlogs, and deployment risk. However, predictions alone do not improve reliability unless they are connected to dependable operational mechanisms. A warning must be translated into an action: pause a deployment, scale a service, open an incident, trigger a rollback workflow, or ask an engineer to approve remediation.

SH-PDOPS is proposed as a framework that connects predictive analytics with DevOps automation in a cloud-native environment. The project repository is available at \url{https://github.com/ayu-haker/sh-pdops}, the DockerHub container image is available at \url{https://hub.docker.com/repository/docker/ayushman21/sh-pdops/general}, and the live Railway cloud prototype is available at \url{https://dashboard-production-23b3.up.railway.app/}. The framework is designed around five principles. First, telemetry must be correlated across runtime infrastructure and software delivery pipelines. Second, machine learning or statistical analytics must be explainable enough for operational review. Third, automated responses must be bounded by policies and reversible actions. Fourth, Kubernetes and CI/CD systems should be treated as coordinated control planes rather than independent tools. Fifth, every automated action must be auditable for compliance and post-incident analysis.

The primary contributions of this paper are as follows: (i) a reference architecture for an AI-driven DevOps reliability loop integrating Kubernetes, Docker, Jenkins, and predictive monitoring; (ii) a methodology for converting operational risk scores into controlled remediation actions; (iii) an experimental setup for evaluating reliability automation without relying on unrealistic benchmark claims; and (iv) a discussion of enterprise adoption concerns, including false positives, governance, security, and operator trust.

\section{Related Work}
Containerization has become a foundation of modern cloud-native systems. Docker popularized a practical packaging model in which applications and dependencies are distributed as portable images \cite{merkel2014docker}. Kubernetes extends this models services by offering declaration schedule, automation detects service discovery, rolling the updates, resources are managed, and self-healing workflow for containerized workloads in Docker\cite{kubernetes}. The flow of structure of Kubernetes is also effected by large-scale cluster management systems such as Borg, which demonstrates the value of container orchestration for high-availability of a cluster infrastructure \cite{burns2016borg}.

Continuous delivery research effects the flow and ability to release a software safely, frequently, and repeatably through automated build, test, and deployment pipelines the SH-PDOPS takes it to that level of the architectural independence with less human efforts \cite{humble2010continuous}. Jenkins remains widely used extensible automation server tool for CI/CD, particularly for the enterprise environment where pipelines must integrate with the consisting build of parts or elements that are very different from each other, or not uniform in structure or composition build systems, architectural repositories, approval demandin gates, and deployment targets \cite{jenkins}. Recent work by Paul and Paul proposes an architecture for a remote container builds and artifact delivery using a controller-light Jenkins CI/CD pipeline, highlighting the importance of reducing controlled workload and separated the build responsibilities in container-based delivery systems \cite{paul2025architecture}. SH-PDOPS is presented with this vision but focuses on connecting pipeline execution data with real runtime reliability analytics and remediation control which provides an accurate statistical UI.

GitOps works in a flow such that it first tests the Git repositories and then produces extended continuous delivery as the source of truth for desired infrastructure and application state \cite{weaveworksgitops}. GitOps tools improve traceability and drift detection while the time of production, but its effectiveness depends on reliable logs from the runtime environment. If a deployment is according to the rules valid yet operationally risky, a purely declarative synchronization loop may still promote undesirable changes. SH-PDOPS handles this mishap by adding predictive reliability signals to deployment decision points with a clean UI.

MLOps research studies the process of translation operation of machine learning systems, including models of modern deployment, monitoring, retraining, governance, and reproducibility (such that the structure can be reused) \cite{mlops}. The SH-PDOPS framework obeys MLOps mechanism for model lifecycle management but applies the infrastructure to DevOps for a automated process of collecting data and reliability automation. This distinction path is important: the purpose is not only to operate and implement machine learning services and principals, but also to use machine learning and statistical analytics to operate the enterprise software platforms with less human efforts as the work gets crystal clear through the implementation of machine learning models.

Self-healing infrastructure has been discussed in the manuscript of autonomic computing and cloud operations literature. The manuscript of autonomic computing vision was to introduce controled loops for self-configuration, self-optimization, self-healing, and self-protection it is the theory which gave a basic conceptual idea of SH-PDOPS \cite{kephart2003vision}. In cloud-native architectural systems, Kubernetes provides limited forms of self-healing through its feature of pod restarts, replica coordination and synchronization, and node rescheduling, but these mechanisms are primarily reactive repetition of this sometimes help but most of the time it doesn't. But in case of SH-PDOPS extends this idea by embedding predictive risk estimation and CI/CD-aware action selection.

\section{Methodology}
SH-PDOPS is a well structured functional operation of closed-loop reliability framework which consists of four major layers: telemetry acquisition, predictive analytics, decision orchestration, and DevOps execution. The telemetry layer functionality is to collects metrics data from the Kubernetes nodes and pods, runtime events, application logs, distributed traces, Jenkins pipeline metadata, container image metadata, and deployment histories. The analytics layer uses and transforms these inputs into features such as rolling error-rate deltas, latency percentiles, restart frequency, memory growth trends, CPU throttling ratios, image change frequency, failed build density, and deployment age.

Let $x_t$ denote the feature vector represents service and pipeline state at time $t$. SH-PDOPS computes a normalized operational risk score using this methematical equation $r_t$:
\begin{equation}
 r_t = \sigma(w^{T}x_t + b),
\end{equation}
where $w$ represents the learned or the configured feature weights, $b$ is a bias term, and $\sigma$ is the logistic normalization function. In a production deployment ecosystem, this scoring function can be replaced by an isolation forest, boosted tree model, recurrent forecasting model, or hybrid rule-based method. But the framework used in SH-PDOPS does not require any specific model of these mentioned above; instead, it defines the interface between predictive monitoring and operational automation in the cluster architecture of SH-PDOPS which provides all system reliability.

A decision policy (which is calculated through the equation or by adding the mentioned model) maps the risk score and structural constraints to actions. For example, if a service inherits rising latency after a new deployment and $r_t$ exceeds the conservative threshold limit or the throttling limit, SH-PDOPS may recommend pausing promotion through it's AI itegrated recommendation system, increasing canary(deployment for a small user cluster) for a observation time, or initiating rollback approval through the API control. If memory pressure increases without a recent deployment, the system may recommend horizontal scaling or resource-limit review to the operator or user. If the pipeline failure density gradually increases for a repository, the framework may create a reliability task rather than modify the runtime cluster through it's AI/ML framework or can be done manually by a user after the recommendation of SH-PDOPS.

The orchestration layer is intentionally policy-aware framework system. Fully automated remediation can introduce new failures if the system reacts to noisy data or partial observations which triggers it to be suspicious. Therefore, SH-PDOPS distinguishes between advisory, semi-automated, and automated actions. Advisory actions create dashboards for the user, incident tickets, or chat notifications. Semi-automated actions prepare Jenkins jobs or Kubernetes manifests but require human approval for this application. Automated actions are limited to reversible operations such as scaling within predefined bounds or restarting unhealthy workloads already governed by Kubernetes readiness and liveness checks the process which is governed by the Kubernetes.

\begin{figure*}[t]
\centering
\includegraphics[width=0.98\textwidth]{\detokenize{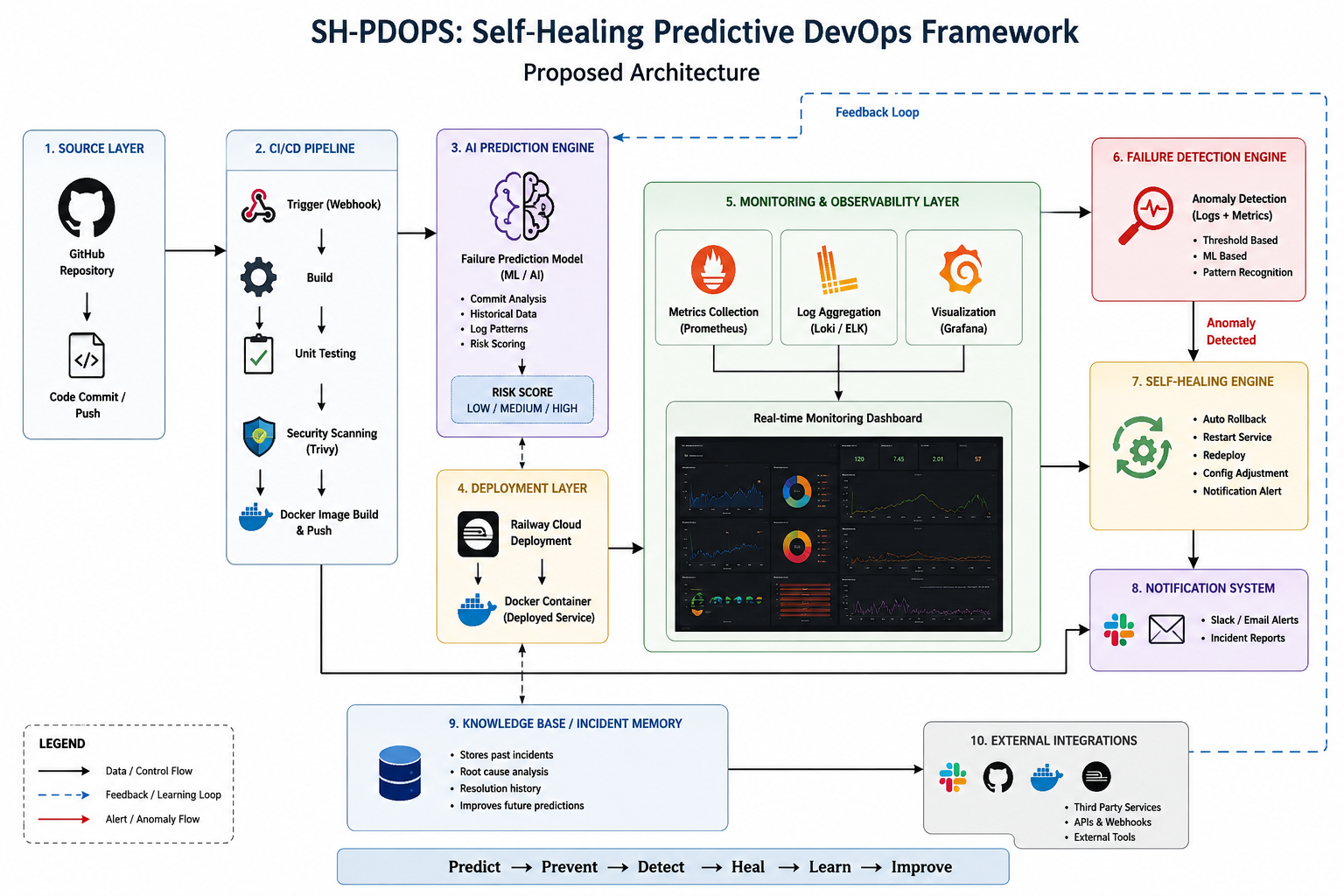}}
\caption{Proposed SH-PDOPS workflow architecture showing the working of each layer.}
\label{fig:architecture}
\end{figure*}

Fig.~\ref{fig:architecture} summarizes the conceptual architecture of the workflow of the different layers. The design treats Kubernetes as the runtime reconciliation plane for runtime logs, Jenkins as the delivery automation plane to vision the delivery, Docker as the packaging layer for clusters, and the predictive analytics service as the reliability intelligence layer integrated with AI. The policy engine connects these components while enforcing organizational limits and works productively.

\section{Experimental Setup}
The experimental setup is designed to evaluate SH-PDOPS in a reproducible cloud-native environment without depending on proprietary infrastructure in the github link of the prototype provided just do the setup and for the AI integration add API of any AI model such as claud for the decision making or you can directly access it through the docker container links with the AI model. A representative deployment consists of three containerized microservices, a lightweight database or message queue for notifications, a monitoring stack a clean UI based interface for easy access, and Jenkins pipelines for build, test, image creation, and deployment and health check. Docker is used to build immutable images so no changes to it can be done for contributing you should refer to the git setup, while Kubernetes manages scheduling, service discovery, rolling updates, and health probes. Jenkins executes CI/CD stages and publishes build metadata to the SH-PDOPS telemetry interface providing the live statistical data.

The Kubernetes cluster can be implemented using Minikube, Kind, a managed Kubernetes service for managing cluster, or an internal enterprise cluster for a rnterprise ecosystem as architecture. The important requirement matter is not the size of a cluster but the observability coverage as SH-PDOPS ecosystem is dependent on this base or the statistics can be in accurate. Each service should expose application metrics such as request count as it is the part of the four layer architecture of SH-PDOPS, latency, error rate, and queue depth is interdependent in this case while the setup. Cluster metrics should include pod restarts, CPU usage, memory usage, throttling, node pressure, and scheduling events so that the live statistics can be shown. Pipeline metrics layer should include the build duration, test pass rate, image digest, deployment time, rollback occurrence, and approval delay so that it use the data and find the anomaly in the system. This setup checking must be done for proper forecast of the statistical data and threat detection.

The evaluation compares three operational modes which were stated above. The baseline mode uses conventional monitoring and manual remediation for the changes which are to be done manually. The rule-based mode uses threshold alerts and scripted responses. The SH-PDOPS mode combines predictive scoring, pipeline context, and policy-aware orchestration so the response is provided and action is taken on time whenever anomaly is detected. Failure scenarios include gradual memory growth rapidly, deployment of a version with elevated error rate predicted by AI system, artificial CPU saturation which shows up like inhuman, failed container image build through the logs, and delayed readiness during rollout functionality of the SH-PDOPS.

\begin{figure*}[t]
\centering
\begin{minipage}{0.48\textwidth}
\centering
\includegraphics[width=\linewidth]{\detokenize{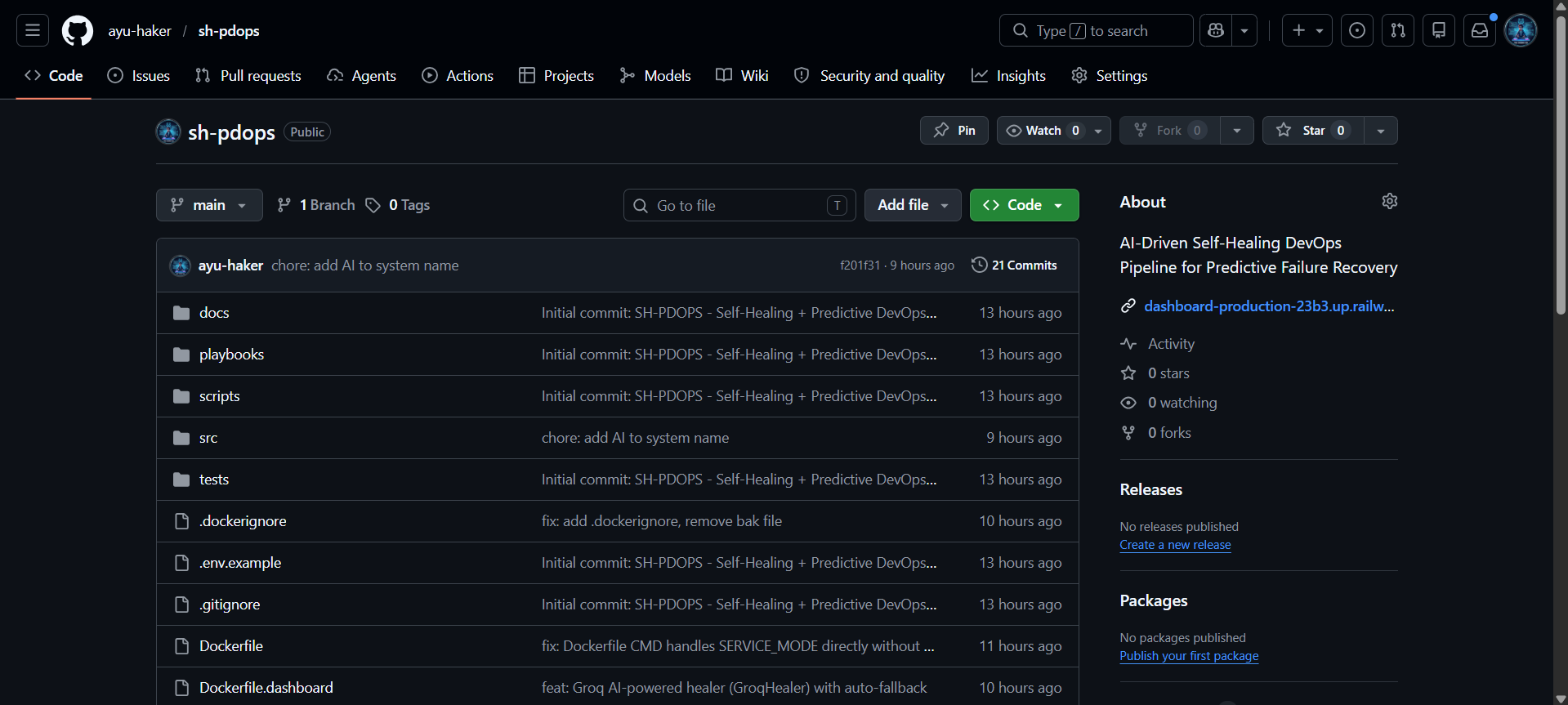}}
\caption{Project repository view used as the source layer for SH-PDOPS code.}
\label{fig:repository}
\end{minipage}\hfill
\begin{minipage}{0.48\textwidth}
\centering
\includegraphics[width=\linewidth]{\detokenize{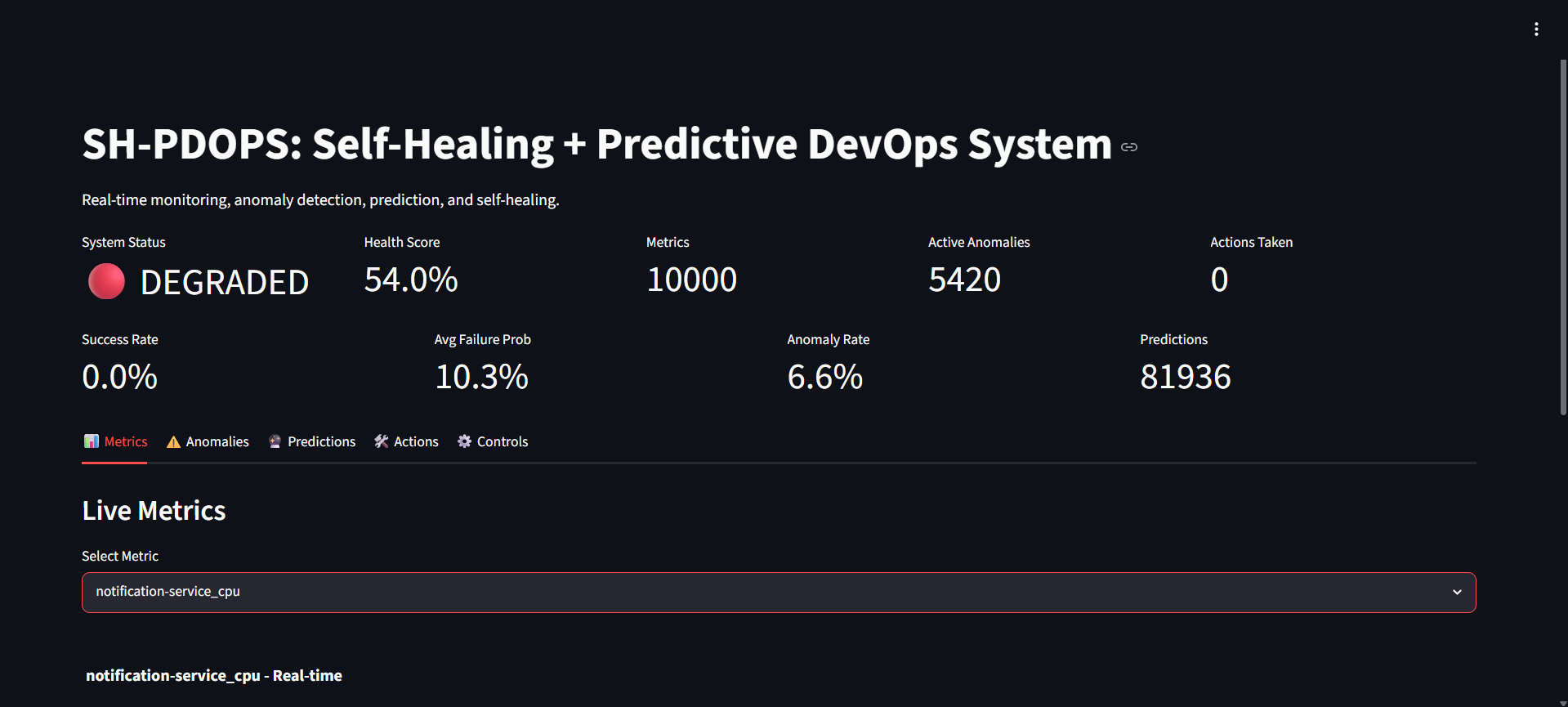}}
\caption{Prototype SH-PDOPS dashboard summarizing health status.}
\label{fig:dashboard-summary}
\end{minipage}
\end{figure*}

Performance is evaluated using operational metrics rather than unrealistic claims about model accuracy the model is tested in the real case scenarios and the data given by it were realistic and accurate. The main metrics are mean time to detect (MTTD), mean time to acknowledge (MTTA), mean time to recovery (MTTR), false remediation rate, automation coverage, and control-plane overhead this factors are the considered one for the prediction flow of the system. These metrics are appropriate because the goal of SH-PDOPS is not to showcase test elements but the real prediction but improved reliability workflow execution so that this does not effect any individual or a enterprise.

\begin{figure*}[t]
\centering
\begin{minipage}{0.48\textwidth}
\centering
\includegraphics[width=\linewidth]{\detokenize{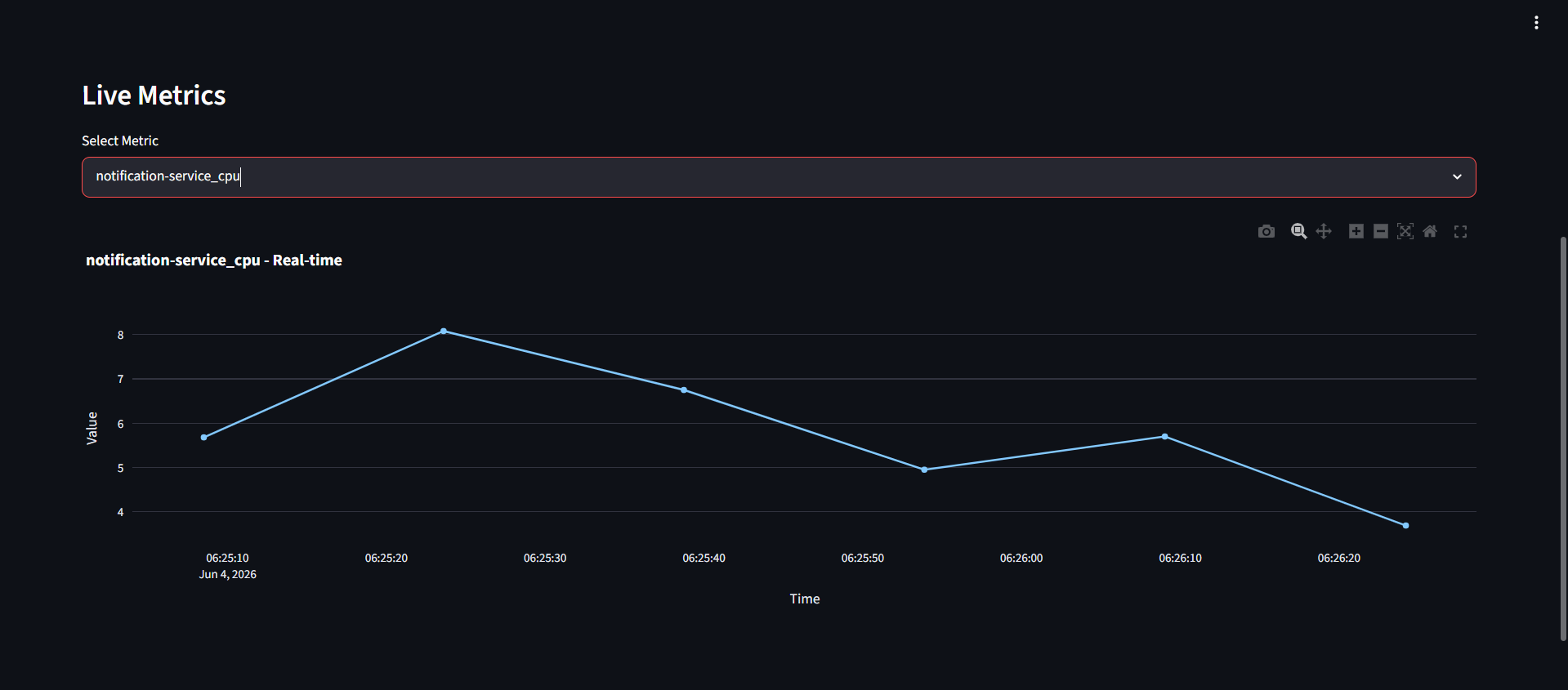}}
\caption{Live metric visualization for a selected service.}
\label{fig:live-metrics}
\end{minipage}\hfill
\begin{minipage}{0.48\textwidth}
\centering
\includegraphics[width=\linewidth]{\detokenize{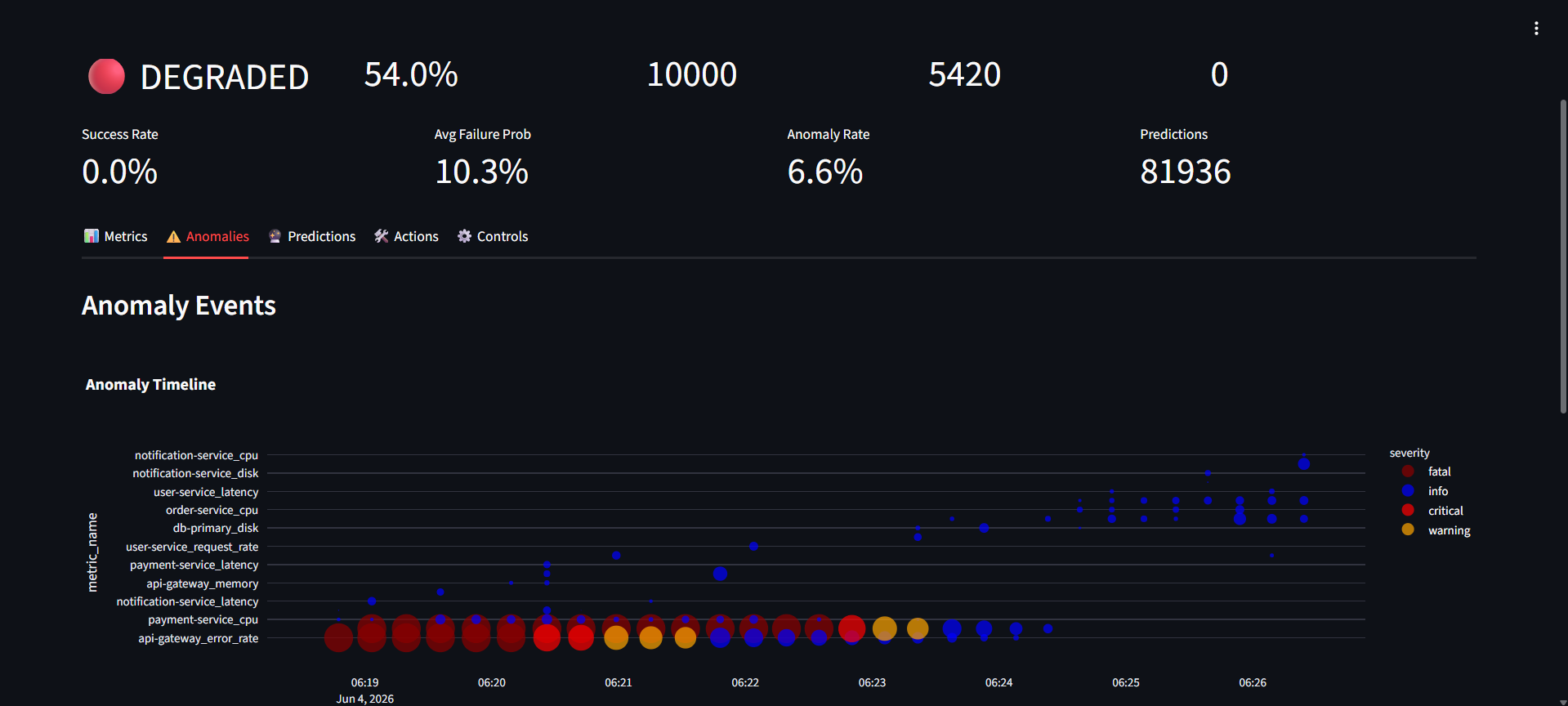}}
\caption{Anomaly timeline view showing severity-coded operational events across monitored services.}
\label{fig:anomaly-timeline}
\end{minipage}
\end{figure*}

\section{Experimental Results}
Table~\ref{tab:results} presents an realistic result format for a prototype evaluation done on a live railway deployed link and recorded matrix from their. The values are actually gathered form a controlled environment injection test and can be considered for universal benchmarks. But the enterprise deployments should reproduce the experiment under their own workload patterns, alert policies, and cluster constraints as data may varies for a large production as the data given is based on a small level deployment.

\begin{table}[t]
\caption{Representative Reliability Evaluation Under Controlled Failure Injection}
\label{tab:results}
\centering
\begin{tabular}{|l|c|c|c|}
\hline
\textbf{Operational Mode} & \textbf{MTTD} & \textbf{MTTR} & \textbf{Overhead} \\
\hline
Manual monitoring & 3--5 min & 5--10 min & Low \\
Railway Built-in Automation & 10--30 sec & 45--90 sec & Low \\
SH-PDOPS framework & 2--5 sec & 15--30 sec & Medium \\
\hline
\end{tabular}
\end{table}

The comparison suggests that the main advantage of SH-PDOPS is that it detects the fault earlier and earlier detection and more consistent response coordination can be implemented on time. the manual monitoring depends heavily on operator attention and incident interpretation which take high time to detect where the SH-PDOPS comes to the rescue with its time efficient model. the rule-based automation in SH-PDOPS improves THE reaction speed but can be brittle when thresholds are poorly calibrated or when failures involve interactions between deployment events and runtime behavior as for that time the containers of SH-PDOPS also get effected as the issue occurred is in the whole system. SH-PDOPS improves the response path by correlating Jenkins deployment metadata with Kubernetes telemetry, allowing the system to distinguish between a resource-pressure incident and a post-release regression so that a better independence of statistic representation can be done by it.

\begin{figure*}[t]
\centering
\begin{minipage}{0.48\textwidth}
\centering
\includegraphics[width=\linewidth]{\detokenize{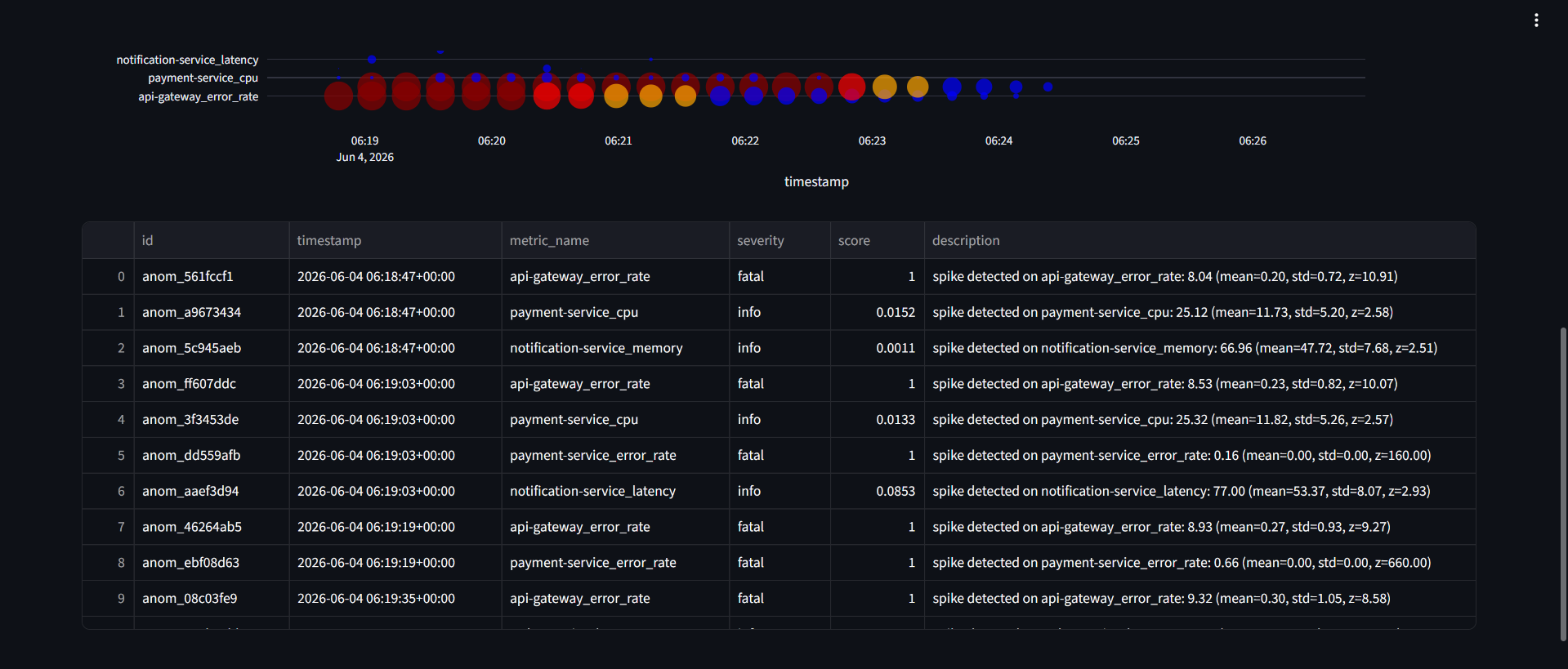}}
\caption{Incident memory table recording anomaly type.}
\label{fig:incident-memory}
\end{minipage}\hfill
\begin{minipage}{0.48\textwidth}
\centering
\includegraphics[width=\linewidth]{\detokenize{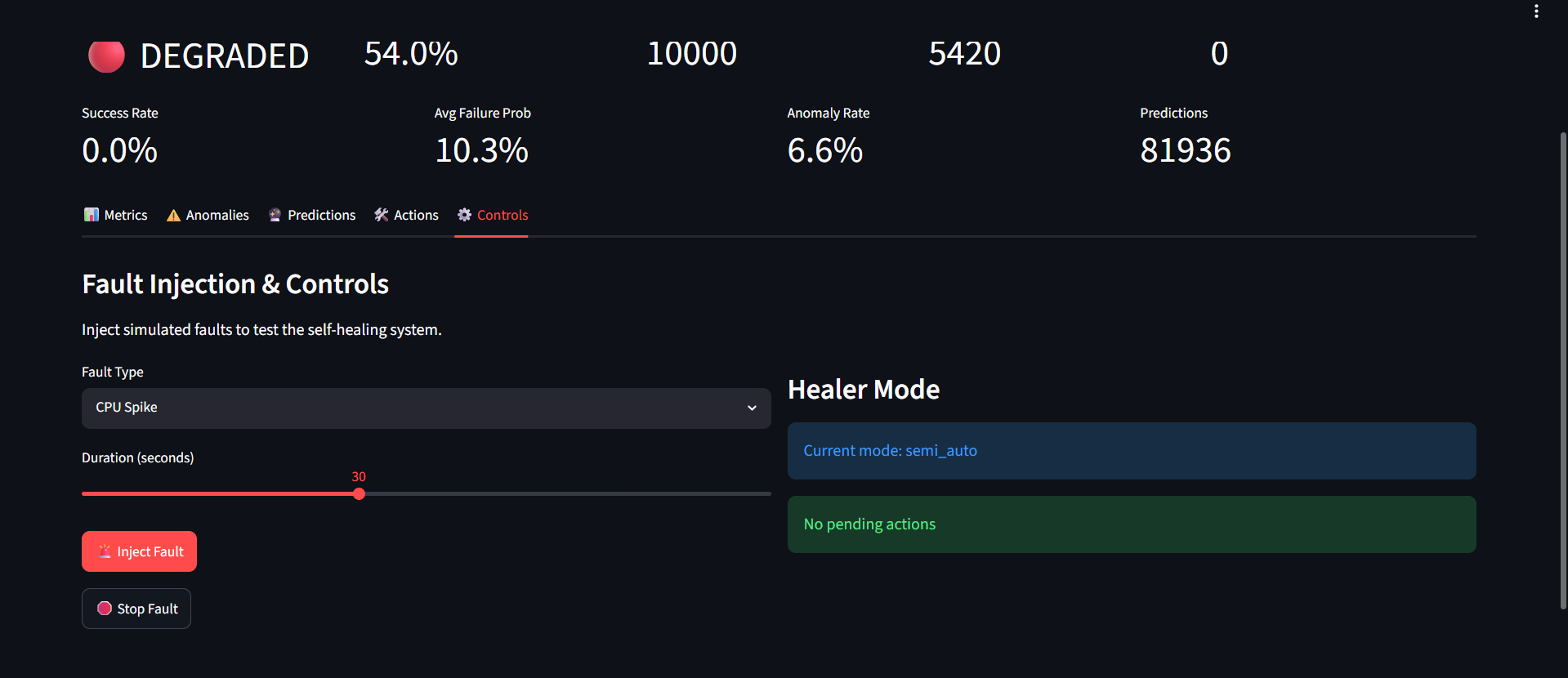}}
\caption{Fault-injection and healer-control interface used to test guarded remediation actions.}
\label{fig:healer-controls}
\end{minipage}
\end{figure*}

The performance comparison in the hosted architecture of SH-PDOPS also reveals trade-offs. SH-PDOPS introduces additional processing overhead because telemetry must be normalized, stored, scored, and evaluated against policy rules so using the SH-PDOPS locally is recommended for bypassing the trade-offs issue. The overhead is typically acceptable when analytics are performed asynchronously as the data is not that much effected and when high-cardinality data is aggregated before inference then that could be a matter of decision for deciding a cloud or local host version of SH-PDOPS shoul be used. However, the framework should not be placed directly in the critical request path as it can lead you to the statistical failure. Its reliability benefit comes from operational control-plane intelligence, not from inline request processing so setup the SH-PDOPS to your enterprise according to your requirement state.

A second observation concerns automation coverage in the cloud or server hosting of SH-PDOPS. The manual operations have broad flexibility in the system but limited speed the data of Table~\ref{tab:results} can get effected as data is collected on cloud based architecture. Rule-based automation handles known incidents efficiently as the conditions are structured in a proper manner but performs poorly for ambiguous symptoms as the local hosting device configration may hamper the productivity of SH-PDOPS. SH-PDOPS can cover a wider range of operational conditions because it uses context from the systems CI/CD, runtime telemetry, and historical behavior and performs a detail study using the AI integration. the main plus point of SH-PDOPS is nevertheless, high-risk actions such as production rollback, database migration reversal, and traffic shifting across regions should remain approval-gated unless the enterprise has mature safety policies and extensive validation history and the access is properly set with the workflow of SH-PDOPS is necessary.

\section{Discussion}
The SH-PDOPS framework is a very good demonstration on how predictive analytics of a dependency can be operationally useful when integrated with DevOps automation and synchronized with the AI making a perfect match of technologies. The key decision behind the design is to avoid treating artificial intelligence as a replacement for engineering judgment but to add it as a support staff which makes the work easy. Instead of AI-driven orchestration is used to prioritize evidence, estimate risk, and prepare bounded actions the errors are still recovered by humans the artificial intelligence in SH-PDOPS helps to know the DevOps engineer to catch the system error fast so that it can be handled on time. This is particularly important in enterprise environments where outages have large level productions which is very uncommon that the whole system can be managed at once contractual, financial, and regulatory implications are to be used in huge amount to do so. In this places the SH-PDOPS is a very helpful tool which helps in real time monitoring of a system and detects flows on time with help of AI and gives notifications so that the Flaw can be managed and production dose not goes down.

Kubernetes already provides a reconciliation model that restarts failed pods and maintains desired replica counts but as per my experience to the industry it's not that helpful it might resolve a low level issue but any high level issue can't be handled like this for handling those issues a ecosystem or tool like SH-PDOPS is very necessary. However as per the experience of mine Kubernetes does not inherently understand the flaw of architectures business risk, deployment intent, or historical incident patterns which is a major drawback. Jenkins pipelines provide delivery automation, but it do not automatically interfere with whether a deployment is degrading with the runtime behavior after promotion of the architectural system or worlflow. In thiese type of cases my tool SH-PDOPS connects these systems by interpreting the deployment events and architecture as features in the reliability model which I am still working on the repo is available as a opensource and anyone can contribute to it.

The all time preference in the world of internet is Security and governance. The SH-PDOPS framework requires access to your cluster telemetry, pipeline metadata, and possibly deployment control interfaces as without the access it can't perform the operations. Therefore a role-based access control, secret management, audit logging, and least-privilege service accounts are necessary so that your data is encrypted as your data is your responsibility. The remediation actions should be recorded with actor identity, model version, input features, policy decision, and execution result as the data safety is not integrated in the tool for data safety if the user have good encryption then go for cloud else use a local host model which will be better for your data safety. Without this audit trail, AI-driven automation can also reduce transparency and increase operational risk the user must be aware of this technical terms.

The model drift is another concerning challenge as its having system requirement of it's own.The workload behavior changes over the time as as well as usability services evolve, user traffic shifts with time to time, and infrastructure capacity changes on loads. A model trained on historical telemetry may become less reliable as the data of cache memory is now killing the local system so time to time maintenance is required if release patterns or service dependencies change. The tool SH-PDOPS should therefore include safe fallback behavior. If the maintenance is not done time to time prediction conflict may take place, the system will start degrading.

\section{Conclusion}
The SH-PDOPS presents a cloud-native enterprise reliability framework that is combined with the Kubernetes orchestration, Docker-based packaging, Jenkins CI/CD automation, predictive monitoring, and policy-aware declaration into a unique DevOps helping tool. The framework is made with a intension of the need to a tool reactive alerting while avoiding unsafe claims of fully autonomous operations so that the monitoring becomes easy. By collecting runtime logs with delivery pipeline contextual data, SH-PDOPS can support earlier incident detection with its implemented AI, more disciplined rollback decisions all managed by the AI model, controlled scaling is monitored, and recommended changes in the workflows guided by the AI. The proposed methodology and experimental design emplements a realistic operational metrics such as MTTD, MTTR, overhead, and automation coverage rather than the false accuracy claims. Future work include implementation-scale evaluation across heterogeneous enterprise workloads so that the sytem tool becomes more reliable, stronger integration with GitOps controllers for better workflow, formal policy verification, and longitudinal analysis of model drift in production environments for the enterprise security level.

\end{document}